\documentclass[final,5p,times,twocolumn,sort&compress]{elsarticle}
\usepackage{amssymb}
\usepackage{lipsum}

\usepackage{graphics}
\usepackage[T1]{fontenc} % if needed
\usepackage{slashed}
\usepackage{mathtools}
\usepackage{xspace}
\usepackage{xcolor}
\usepackage{amsmath}
\usepackage[symbol]{footmisc}

\newcommand{\gev}{\ensuremath{\,\text{GeV}}\xspace}

\journal{Nuclear Physics B}

\begin{document}

\begin{frontmatter}

\title{New vacuum stability limit from cosmological history}

\author[a]{Csaba Bal\'azs$^1$}
\author[b,c]{Yang Xiao$^2$}
\author[b,c]{Jin Min Yang$^3$}
\author[d]{Yang Zhang$^4$}

\affiliation[a]{organization={School of Physics and Astronomy, Monash University},%Department and Organization
            addressline={}, 
            city={Melbourne},
            postcode={3800}, 
            state={Victoria},
            country={Australia}}
\affiliation[b]{organization={CAS Key Laboratory of Theoretical Physics, Institute of Theoretical Physics, Chinese Academy of Sciences,},%Department and Organization
            addressline={}, 
            city={Beijing},
            postcode={100190}, 
            state={},
            country={P. R. China}}
\affiliation[c]{organization={School of Physical Sciences, University of Chinese Academy of Sciences},%Department and Organization
            addressline={}, 
            city={Beijing},
            postcode={100049}, 
            state={},
            country={P. R. China}}
\affiliation[d]{organization={School of Physics, Zhengzhou University},%Department and Organization
            addressline={}, 
            city={Zhengzhou},
            postcode={450001}, 
            state={Henan},
            country={P. R. China}}

\begin{abstract}

The stability of the electroweak vacuum imposes important 
constraints on new physics models. Such new physics models may 
introduce one or more new thermal phases with a lower free energy than 
that of the electroweak vacuum. In this case, the early universe may stay or
have already evolved into one of these deeper vacuum states. We 
investigate this possibility in detail in the singlet extension of the 
Standard Model, and delineate the corresponding constraints in its 
parameter space. We also discuss the situation in supersymmetry as 
another example. To account for the possibility that the universe is 
trapped in a non-electroweak vacuum, we propose a procedure of calculating 
the vacuum stability limit efficiently.

\end{abstract}


\end{frontmatter}

%\tableofcontents

%% \linenumbers

%% main text

\footnotetext[1]{csaba.balazs@monash.edu}
\footnotetext[2]{xiaoyang@itp.ac.cn, corresponding author}
\footnotetext[3]{jmyang@itp.ac.cn}
\footnotetext[4]{zhangyangphy@zzu.edu.cn, corresponding author}

\section{Introduction}
\label{sec:intro}

Long before the discovery of the Higgs boson in 2012, vacuum decay had been used as a theoretical tool to constrain the masses of the Standard Model (SM) particles~\cite{isidori2001metastability,casas1995improved,espinosa1995improved,altarelli1994lower,sher1993precise,Branchina:2013jra,Branchina:2014usa,Branchina:2014rva}. The critical reason for vacuum instability in the SM is that with loop corrections resummed via the renormalization group, the quartic coefficient of the effective potential becomes negative as the renormalization scale approaches about $10^{10}$ \gev \cite{buttazzo2013investigating}. This means that the effective potential must be ultraviolet completed such that it is bounded from below and a second, deeper minimum arises~\cite{sher1989electroweak,krive1976vacuum}. Fortunately, detailed calculation shows that the tunneling time of the desired electroweak symmetry breaking~(EW-breaking) vacuum into this deeper vacuum is longer than the age of the universe~\cite{buttazzo2013investigating,andreassen2018scale}.

In beyond the SM (BSM) scenarios, such as supersymmetric models, extra scalar fields are usually introduced, which can get non-zero vacuum expectation values and destabilize the EW-breaking vacuum. If another lower energy minimum exists, the EW-breaking vacuum could transit into such a vacuum via quantum tunneling or thermal excitation. Since the EW-breaking vacuum is compatible with experimental results, 
we naturally demand that the transition time is longer than the age of the universe, which constrains the model to describe Nature. In simple models such constraints for stability at zero-temperature can be derived analytically~\cite{ferreira2004stability,barroso2007neutral,barroso2013metastability}. To deal with multi-fields or high temperature effects, several numerical tools were developed~\cite{camargo2013vevacious,masoumi2017efficient,sato2021simplebounce,Hollik:2018wrr,Ferreira:2019iqb}. 
% to assess vacuum stability we need to resort to numerical calculation tools as \textbf{Vevacious}~\cite{camargo2013vevacious}. 
Both analytic and numerical calculations assume that the set of considered parameters ensure the realization of the EW-breaking vacuum during the evolution of the universe. 

Examining the cosmological history, however, reveals that the above assumption is not always correct. 
The early universe phase transition associated with the EW symmetry breaking has gained a lot of attention because it can lead to baryogenesis and a potentially detectable stochastic gravitational wave background~\cite{Pietroni:1992in,Cline:1996mga,Ham:2004nv,Funakubo:2005pu,Barger:2008jx,Chung:2010cd,Espinosa:2011ax,Chowdhury:2011ga,Gil:2012ya,Carena:2012np,No:2013wsa,Dorsch:2013wja,Curtin:2014jma,Huang:2014ifa,Profumo:2014opa,Kozaczuk:2014kva,Jiang:2015cwa,Curtin:2016urg,Vaskonen:2016yiu,Dorsch:2016nrg,Huang:2016cjm,Chala:2016ykx,Basler:2016obg,Beniwal:2017eik,Bernon:2017jgv,Kurup:2017dzf,Andersen:2017ika,Chiang:2017nmu,Dorsch:2017nza,Beniwal:2018hyi,Bruggisser:2018mrt,Ellis:2018mja,Athron:2019teq,Kainulainen:2019kyp,Bian:2019kmg,Li:2019tfd,Chiang:2019oms,Han:2020ekm,Xie:2020bkl,Bell:2020gug,Lewicki:2021pgr,Wang:2022yhm,Azatov:2022tii,Xiao:2022oaq}.  It has long been known that the EW phase transitions can not always successfully proceed.
Cline \emph{et al.}~\cite{cline1999electroweak} investigated whether the transition process from a color-breaking vacuum to the EW-breaking vacuum can happen and found that the nucleation rate is many orders of magnitude too small even if the parameters are optimized. Recently, Baum \emph{et al.}~\cite{baum2021nucleation} stressed the importance of nucleation temperature in determining whether the tunneling process can happen and also pointed out that the two-step phase transition will cease due to lack of nucleation temperature and hence the vacuum will be trapped at the trivial minimum. Meanwhile, Biek{\"o}tter \emph{et al.} also found that in the 2HDM~\cite{2hdmtrapped}, N2HDM~\cite{biekotter2021fate} and $\mu\nu$SSM~\cite{Biekotter:2021rak} the transition from the electroweak symmetry restored (EW-restored) minimum to the desired EW-breaking vacuum may not be achieved. In these papers, this limit is considered a necessary condition for a successful EW phase transition.

In this paper, we demonstrate that, even when not considering baryogenesis or gravitational waves, in order to obtain proper constraints from the EW vacuum stability requirement, it is necessary to check the evolution history.
We show cases when, in principle, it is possible to transition to the EW vacuum but this transition does not actually happen.  We also find cases when a non-EW vacuum is always the global minimum during the thermal evolution, which also have to be excluded.
As a result, the full thermal history plays an important role in addition to the zero or low-temperature analysis of vacuum stability.

To demonstrate the above scenarios, we use the $Z_{2}$ symmetric real scalar extension of the SM (xSM).  The EW phase transition has been extensively studied in the context of the xSM (see, for example, references \cite{Espinosa:2011ax,Profumo:2014opa,Curtin:2014jma,Beniwal:2017eik,Kurup:2017dzf,Ellis:2018mja,Lewicki:2021pgr,Azatov:2022tii}).  We reveal that the parameter space corresponding to long-lived meta-stable EW-breaking vacuum and some of the parameter space corresponding to stable EW-breaking vacuum are excluded due to the requirement that cosmological evolution lands the universe in the EW vacuum at zero temperature.
Additionally, we find that checking the cosmological history can be computationally advantageous in certain regions of the xSM parameter space.

This paper is organized as follows.  In Section 2, we address in detail the importance of checking the evolution history of the universe in the study of EW vacuum stability.  Section 3 describes our toy model and calculations.  Section 4 compares the limits obtained from the traditional method and from checking the cosmological history.  Finally, we summarize our findings in Section 5.

\begin{figure*}[t]
\centering 
\includegraphics[width=\textwidth]{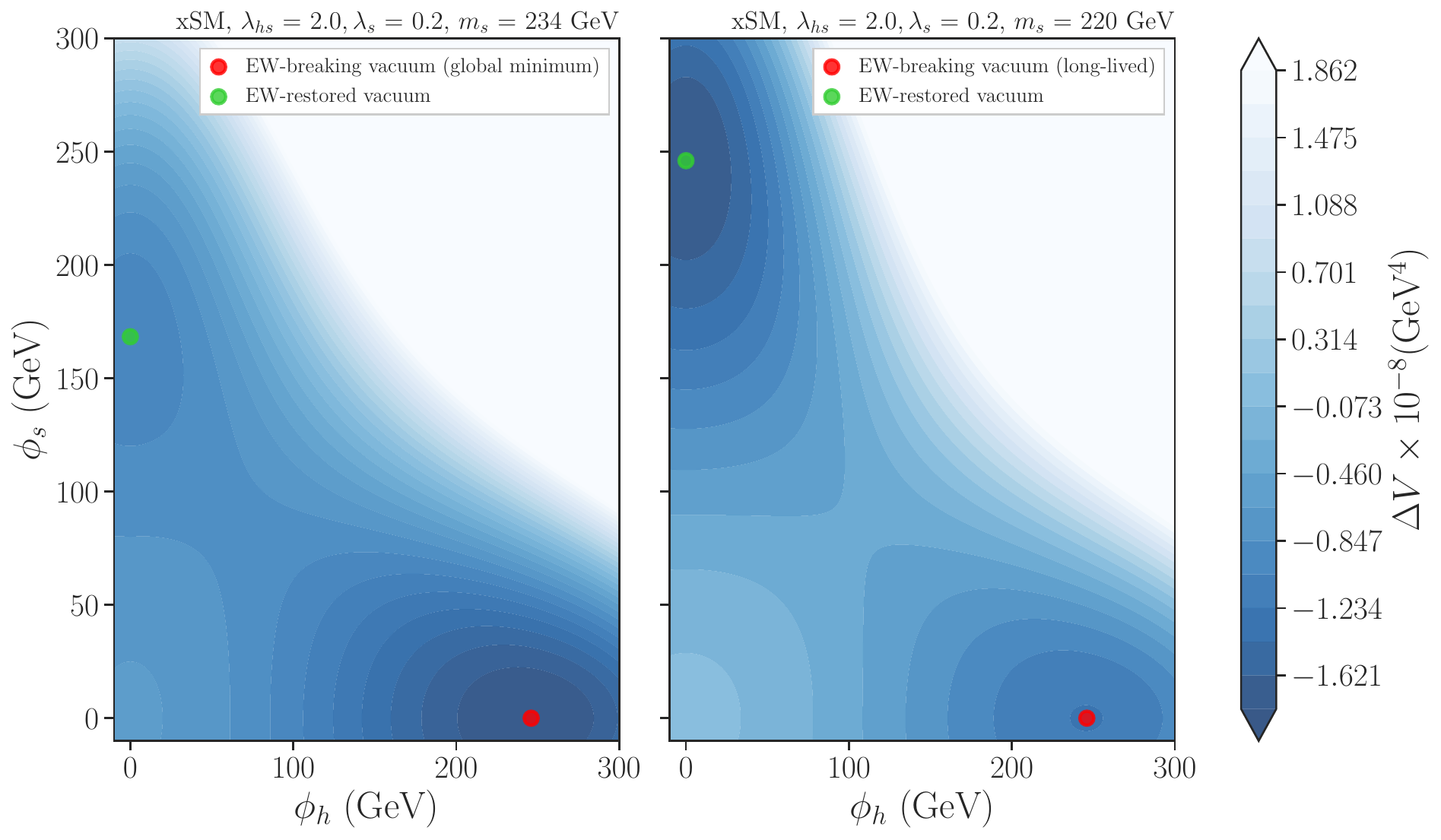}
\caption{ Two situations in which the desired EW-breaking vacuum is stable (left panel) and long-lived (right panel). The values of the SM-like Higgs field $\phi_h$ and the singlet field $\phi_s$ are shown on the $x$ and $y$ axes, respectively. The color scale indicates the (scaled) value of the effective potential: $\Delta V = V(h,s)-V(0,0)$. } 
\label{fig:bk0}
\end{figure*}

\section{Vacuum stability analysis with cosmological history}
\label{sec:mechanism}

Based on the (near-)zero temperature effective potential of a new physics model, the traditional vacuum stability analysis considers two scenarios.
In Fig.~\ref{fig:bk0} we illustrate these two scenarios using the effective potential of the xSM plotted in the plane of the SM-like Higgs field $\phi_h$ and the extra singlet field $\phi_s$.  (We describe the xSM model later in detail.)  Color indicates the corresponding value of the potential, the red dot, at $\phi_h =246~\gev$, represents the desired EW-breaking vacuum that agrees with experimental results, and the green dot corresponds to another vacuum.
In the left frame, the desired EW-breaking vacuum at zero-temperature is a global minimum, making it completely stable.  A meta-stable EW-vacuum is shown in the right frame.  If the tunneling time from it to the global minimum, indicated by a green dot, is longer than the age of the universe, the corresponding point in the parameter space is deemed viable.  The traditional vacuum stability analysis accepts both of these scenarios as valid cases.

\begin{figure*}[t]
\centering 
\includegraphics[width=0.49\textwidth]{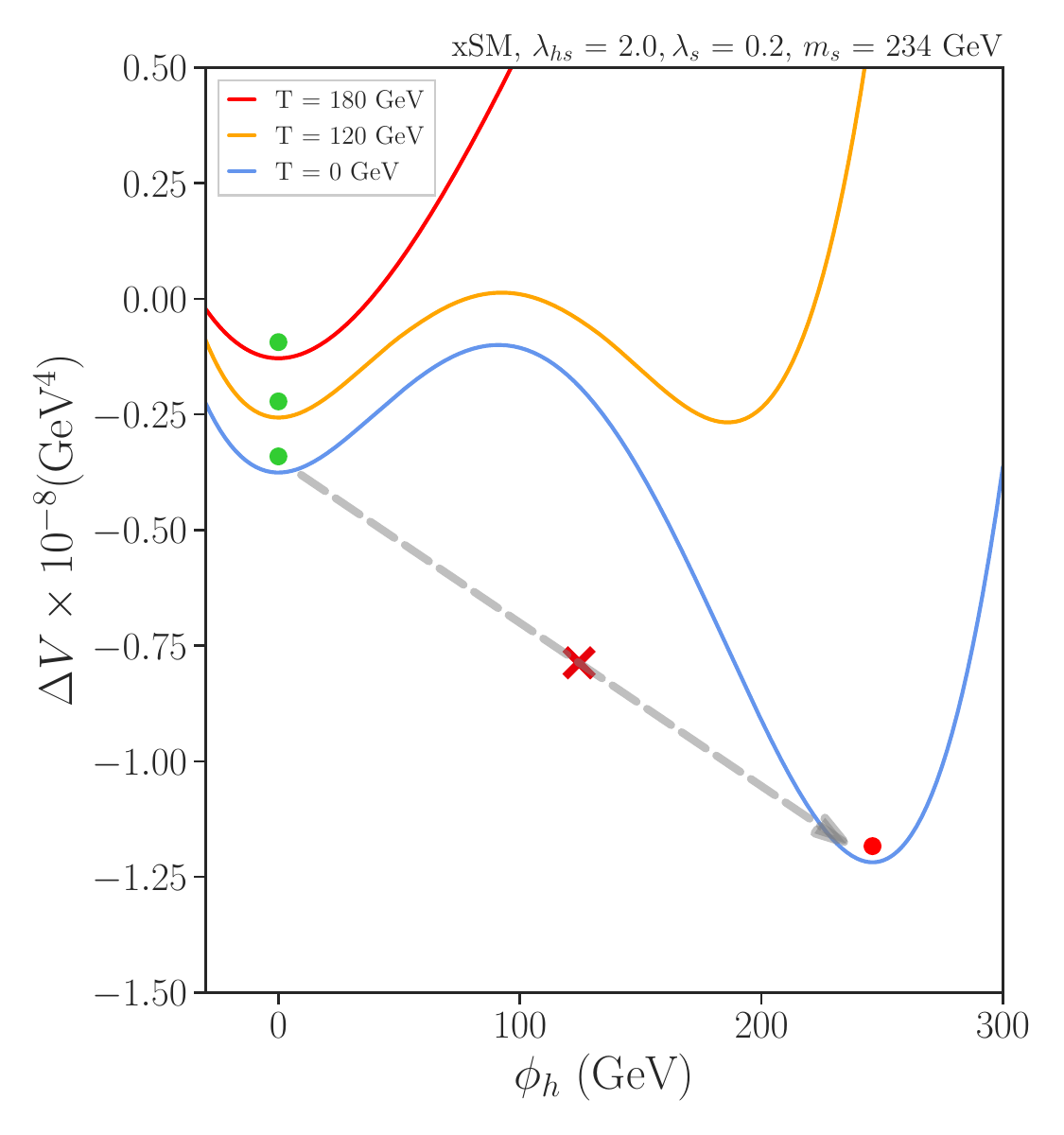}
\includegraphics[width=0.49\textwidth]{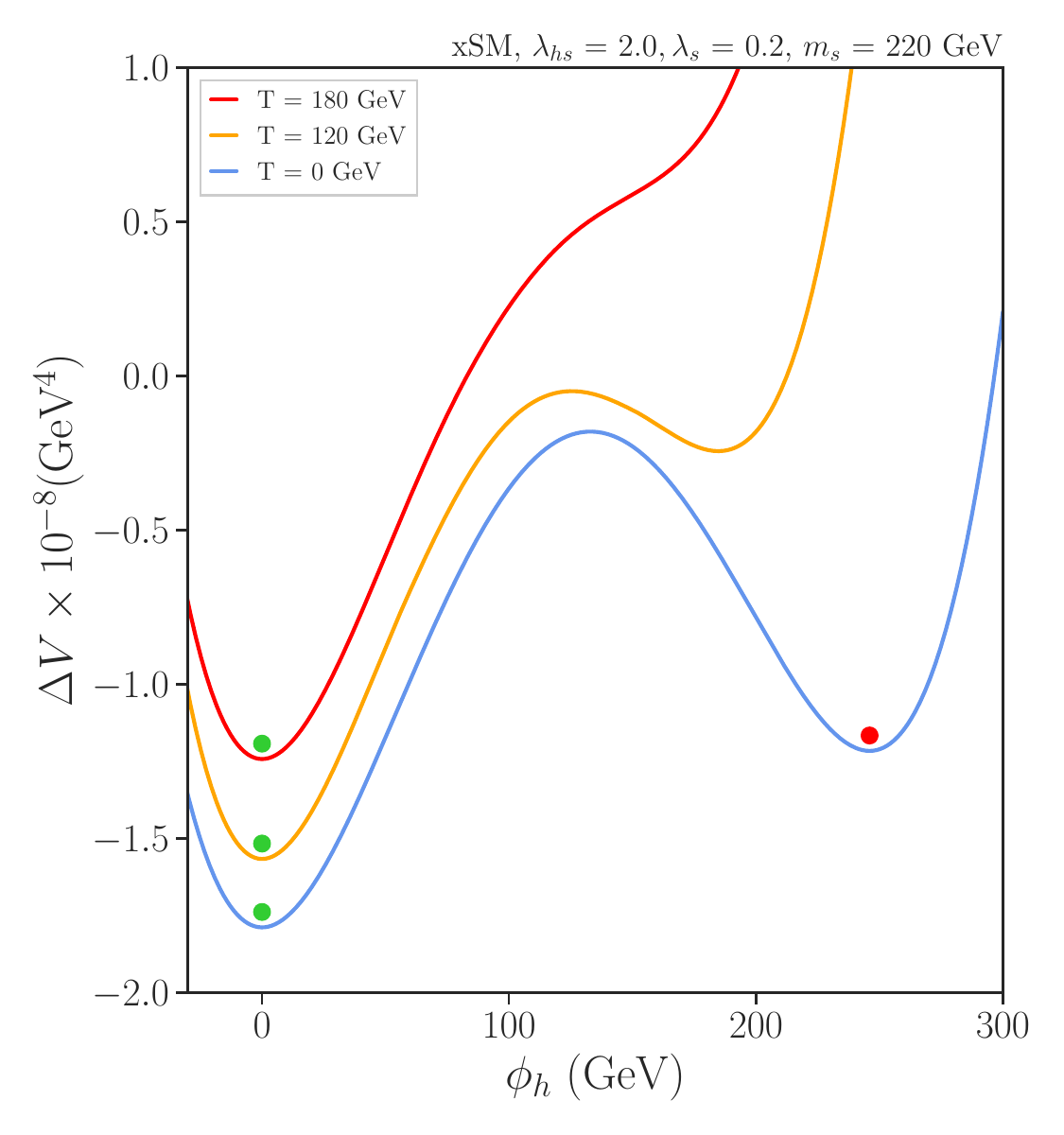}
\caption{In the scenarios shown in Fig.~\ref{fig:bk0} the universe always stays in the non-EW vacuum during the thermal history.  The solid green ball indicates the vacuum where the universe stays at the corresponding temperature, while the red ball stands for the SM EW-breaking vacuum at zero-temperature.  The arrow with cross means that the transition can not take place.  These scenarios can be realized in the xSM using parameters shown by the labels.  The scaled effective potential is along the direct path between the two minima $(0,v_s)$ and $(v_h,0)$.}  
\label{fig:bk1}
\end{figure*}

We find, however, that this analysis is insufficient. In Fig.~\ref{fig:bk1} we present the effective potential at finite temperatures for the exactly same benchmarks as in Fig.~\ref{fig:bk0}.  The potentials are displayed in one dimension along the direct path connecting the two minima (red and green dots in Fig.~\ref{fig:bk0}).  In these scenarios the EW-breaking vacuum is stable or long-lived at zero temperature but the universe is trapped in a non-EW vacuum.
In the early hot universe, in the global minimum of effective potential (shown by the green dot and red lines) the EW symmetry is restored.  With decreasing temperature a new minimum appears in the potential (shown by the orange lines).  In the left frame, the new minimum turns into the desired EW-breaking vacuum and global minimum at zero-temperature, but the transition from the EW-restored vacuum to it may not happen\footnote{Note that the EW-restored vacuum in the xSM may have non-zero vacuum expectation value for $\phi_s$.}.  In the right frame, the new minimum also evolves into the desired EW-breaking vacuum but always has higher free energy than that of the EW-restored vacuum, so there is no transition at all. Therefore, the correct way to check the vacuum of today's universe to follow the thermal history, starting from high temperature, calculating every possible transition and tracing the minima of the potential as they evolve to zero-temperature.

In Ref.~\cite{camargo2013vevacious} thermal transitions are taken into account in the study of vacuum stability, but in the traditional view. 
The authors calculate the lifetime of the EW-breaking vacuum at finite temperature, starting from low temperature and ending at the temperature at which one of the vacua evaporates.  In the right frame of Fig.~\ref{fig:bk1}, with  increasing temperature, the transition probability of the desired EW-breaking vacuum increases, which makes it a short-lived vacuum.
Thus, this method can also correctly exclude this situation. However, if there are more than one minima that the desired EW-breaking vacuum can transition into at finite temperature, this method becomes invalid.  As shown later, the calculation of transition probability is fairly time consuming, while checking the thermal history only needs to trace the free energies of the minima for this situation, which precedes calculating transition probability. 

\section{Model and Calculation}

The scenarios displayed in Fig.~\ref{fig:bk0} and Fig.~\ref{fig:bk1} can be realized in the xSM.  We choose this simple model to illustrate that these scenarios may occur in a wide range of BSM models.
The tree-level effective potential of the xSM takes the form
\begin{equation}\label{eq:xsm_tree}
    V_{0}(\phi_h,\phi_s) = -\frac{\mu_{h}^{2}}{2} \phi_h^2 + \frac{\lambda_{h}}{4} \phi_h^4 - \frac{\mu_{s}^2}{2} \phi_s^2 + \frac{\lambda_{s}}{4} \phi_s^4 + \frac{\lambda_{hs}}{4} \phi_h^2\phi_s^2,
\end{equation}
where $\phi_h$ and $\phi_s$ are the physical Higgs and extra scalar fields, respectively.  For simplicity, 
%To keep the zero-temperature loop-corrected minima the same as the tree level, 
we use the on-shell renormalization scheme to include the one-loop corrections~\cite{quiros1998finite}
\begin{eqnarray}\label{eq: CW potential}
V_{1}(\phi_h, \phi_s) = && \sum_{i} (-1)^{s_{i}} \frac{g_{i}}{64 \pi^2}  \left\{  m_{i}^4(\phi_h, \phi_s)\left[\log \frac{m_{i}^2(\phi_h, \phi_s)}{m_{i}^2(v_{h}, v_{s})}  \right.\right.\nonumber\\ 
&&  \left. -\frac{3}{2} \right] + 2m_{i}^2(\phi_h, \phi_s)m_{i}^2(v_{h}, v_{s}) \biggr\},
\end{eqnarray}
where the sum is over $i \in \{h,~S,~W^{\pm},~Z,~\gamma,~t$\}, $g_i$ counts the degrees of freedom of a given field, $m_{i}(\phi_h, \phi_s)$ is the field-dependent mass, $s_i$ is the spin of field $i$, and $v_{h}$ and $v_{s}$ are the expectation values in the EW vacuum at zero-temperature~\cite{anderson1992electroweak}.  The loop contributions from leptons and light quarks are ignored due to their small couplings to the Higgs.  Contributions from Goldstone boson are not considered as they cause divergences in the second derivatives of the potential.  These simplifications affect the limits obtained from checking the zero-temperature potential and from the cosmological history the same way, so they do not change our conclusion.

The one-loop finite temperature correction is deduced from finite-temperature field theory~\cite{quiros1998finite}
\small
\begin{eqnarray} \label{eq: one-loop effective potential at finite temperature}
    V_{1T}(\phi_h, \phi_s; T) =& \frac{T^{4}}{2 \pi^2} \left[ \sum_{B} g_{B}J_{B}\left(\frac{m_{B}^2(\phi_h, \phi_s)}{T^2}\right) \right.\nonumber\\
     & \left. + \sum_{F} g_{F}J_{F}\left(\frac{m_{F}^2(\phi_h, \phi_s)}{T^2}\right)\right],
\end{eqnarray}
\normalsize
where $J_{B}$ and $J_{F}$ are the relevant thermal distribution functions for the bosonic and fermionic contributions, respectively,
\begin{equation} \label{eq: J_BJ_F}
    J_{B,F}(x^2) = \pm \int_{0}^{+\infty} {\rm d}y ~y^2 \log \left(1\mp e^{-\sqrt{x^2 + y^2}}\right) ,
\end{equation}
and $g_{B,S}$ is the number of degrees of freedom.
An estimation made by Weinberg indicates that higher order correction may overwhelm $V_{1T}$ at high temperature and thus a re-summation is required~\cite{weinberg1974gauge, parwani1992resummation, senaha2020symmetry, arnold1993effective}.  We adopt the re-summation framework proposed in \cite{parwani1992resummation}, which replaces the tree-level masses $m_{i}^2$ by the thermal masses $m_{i}^2(T) = m_{i}^2 + d_{i}T^2$, with $d_{i}T^2$ being the leading contribution in temperature to the one-loop thermal masses~\cite{carena2020electroweak}. 
The total effective potential $V$ is given by
\begin{equation} 
    V(\phi_h, \phi_s; T) = V_{0}(\phi_h, \phi_s) + V_{1}(\phi_h, \phi_s) + V_{1T}(\phi_h, \phi_s; T).
\end{equation}
% Typically, to study the vacuum transition in the new physics model, renormalization group (RG) running is used to adjust the parameters just as in the previous study of the SM~\cite{devoto2022false}. However, we here do not adopt this method and just regard the parameters as free ones. The reason for this is that the on-shell scheme is scale independent and we just want to get the constraints on the parameters instead of the corresponding scale. 

Having the effective potential, we can calculate the lifetime\footnote{Here, the lifetime refers to the decay time from the desired EW-breaking vacuum to another vacuum, which differs from the typically used vacuum  lifetime in the SM—i.e., the decay time to a field of infinity. New physics can also influence the decay lifetime towards a field of infinity~\cite{Bentivegna:2017qry,Branchina:2018qlf,Branchina:2018xdh,Branchina:2019tyy}, which is beyond the scope of this article. } and transition probability of the desired EW-breaking vacuum in both the present universe and the early universe. Here we calculate the transition probability following \texttt{Vevacious}~\cite{camargo2014constraining}.
The tunneling rate, per unit time and unit volume, of the meta-stable vacuum at zero-temperature was first calculated by Coleman~\cite{coleman1977fate}
\begin{equation}
    \frac{\Gamma}{V} = A^4e^{-S_{E}} .
\end{equation}
Here $S_{E}$ denotes the Euclidean bounce action which gives the dominant contribution to the tunneling rate, and $A$ is a dimensionful parameter which is hard to calculate and is often roughly estimated as the typical energy scale (such as 170 \gev in our model)~\cite{hollik2019impact}. To determine if the meta-stable vacuum survives to this day, we define the ratio of tunneling time $t_{tun}$ to the universe age $t_{uni}$ like Ref.~\cite{hollik2019impact} 
\begin{equation} \label{eq:tranditional method}
    \frac{t_{tun}}{t_{uni}} = \left(\frac{\Gamma}{V}\right)^{-\frac{1}{4}} \frac{1}{t_{uni}} = \frac{e^{\frac{S_{E}}{4}}}{At_{uni}}.
\end{equation}
If the ratio is greater than one, then the zero-temperature meta-stable vacuum is long-lived, otherwise it is short-lived.

Similarly, the transition rate, per unit time per unit volume, of the meta-stable vacuum at finite temperature can be expressed as~\cite{linde1983decay}
\begin{equation}
    \frac{\Gamma}{V} = B(T)e^{-S_{3}(T)/T} ,
\end{equation}
where $S_{3}$ denotes the Euclidean bounce action over three dimensions and $B(T)$ is a dimensionful parameter, which is often estimated as $T^4$.
%There are two popular ways to calculate it. 
The probability $P(T_i, T_f)$ of non-transitioning (surviving) during the period from an initial temperature $T_i$ to a final temperature $T_f$ can be obtained by trading time $t$ with temperature $T$ in the differential equation
\begin{equation}
    P(t+{\rm d}t) = P(t)(1-\Gamma {\rm d}t).
\end{equation}
As a result, we obtain 
\small
\begin{equation}
    P(T_i, T_f) = \mathrm{exp}\left[-\int^{T_{f}}_{T_{i}}\frac{{\rm d}t}{{\rm d}T}V(T)B(T)e^{-S_{3}(T)/T}{\rm d}T\right],
\end{equation}
\normalsize
where $V(T)$ is the observable volume. 
Assuming the universe is radiation dominated during the evolution from $T_i$ to $T_f$, entropy is approximately conserved and $S_{3}(T)$ increase monotonically with $T$ increasing. The upper bound of $P(T_i, T_f)$ is then expressed as 
\small
\begin{eqnarray} \label{eq:vevacious_P}
    P(T_i = T, T_f = 0) \leq  \mathrm{exp}(-1.581 \times 10^{106} e^{-S_{3}/T}/S_{3}).
\end{eqnarray}
\normalsize
A temperature $T_{opt}$ to maximize the right-hand side of Eq.~(\ref{eq:vevacious_P}) can be found by fitting the thermal action to $b/(T-T_c)^2$, where $T_{c}$ is the critical temperature at which the two minima have degenerate free energy. \texttt{CosmoTrans
tions}~\cite{wainwright2012cosmotransitions} is used to calculate the action. Then we can set a threshold, $20\%$ in this work, on the maximal surviving probability to judge whether the tunneling process occurs or not. Our results are not particularly sensitive to the value of the threshold. 

Previous studies~\cite{baum2021nucleation, 2hdmtrapped, biekotter2021fate} adopted the existence of nucleation temperature as a criterion. 
The transition process is identified with the nucleation process, in which the surface tension of the vacuum bubble wall and the energy of the bubble expansion compete against each other. As the universe cools, the thermal transition rate increases dramatically. The transition can be regarded as happening once the probability to nucleate one critical bubble in the Hubble volume is of order one, at the so-called nucleation temperature $T_n$, which can be roughly estimated from $S(T_{n})/T_{n} \sim 140$ \gev. However, the nucleation temperature may be ill-defined for slow or supercooled transitions~\cite{cai2017gravitational, Athron:2022mmm, Xiao:2023dbb}.

To describe the phase transition more accurately, we could use the fraction of the false vacuum that accounts for the bubble dynamics,
\begin{equation} \label{eq: P(t)}
    P(t) = \exp \left[ -\int^t _{t_{initial}}B(t')e^{-S(t')}V(t',t)dt' \right], 
\end{equation}
where $S(t')$ is $S_3(T)/T$ but the temperature is replaced by time, $V(t',t) = g (\int ^t _{t'} u(\tau) d\tau)^3$, and $u$ is the bubble wall velocity.  The shape constant $g$ equals to $4\pi/3$ for a spherical bubble.
The bubble wall velocity could be obtained by solving the fluid-scalar coupled Boltzmann equation which is well beyond the scope of this paper.  We could regard the bubble wall velocity as a free parameter, but it may introduce a dependence of the exclusion limit on this additional parameter.  Moreover, calculating Eq.~(\ref{eq: P(t)}) is much more time consuming than Eq.~(\ref{eq:vevacious_P}). 

We have verified that the limits for having a successful EW phase transition obtained from the above three approaches are similar to each other, and all of them are more stringent than the limits for having a long-lived or stable EW vacuum at zero-temperature.  Therefore, we use Eq.~(\ref{eq:vevacious_P}), which gives the probability of non-transitioning, as the transition criterion in the following analysis.

%%%fig.3 
\begin{figure*}[t]
\centering 
\includegraphics[width=\textwidth]{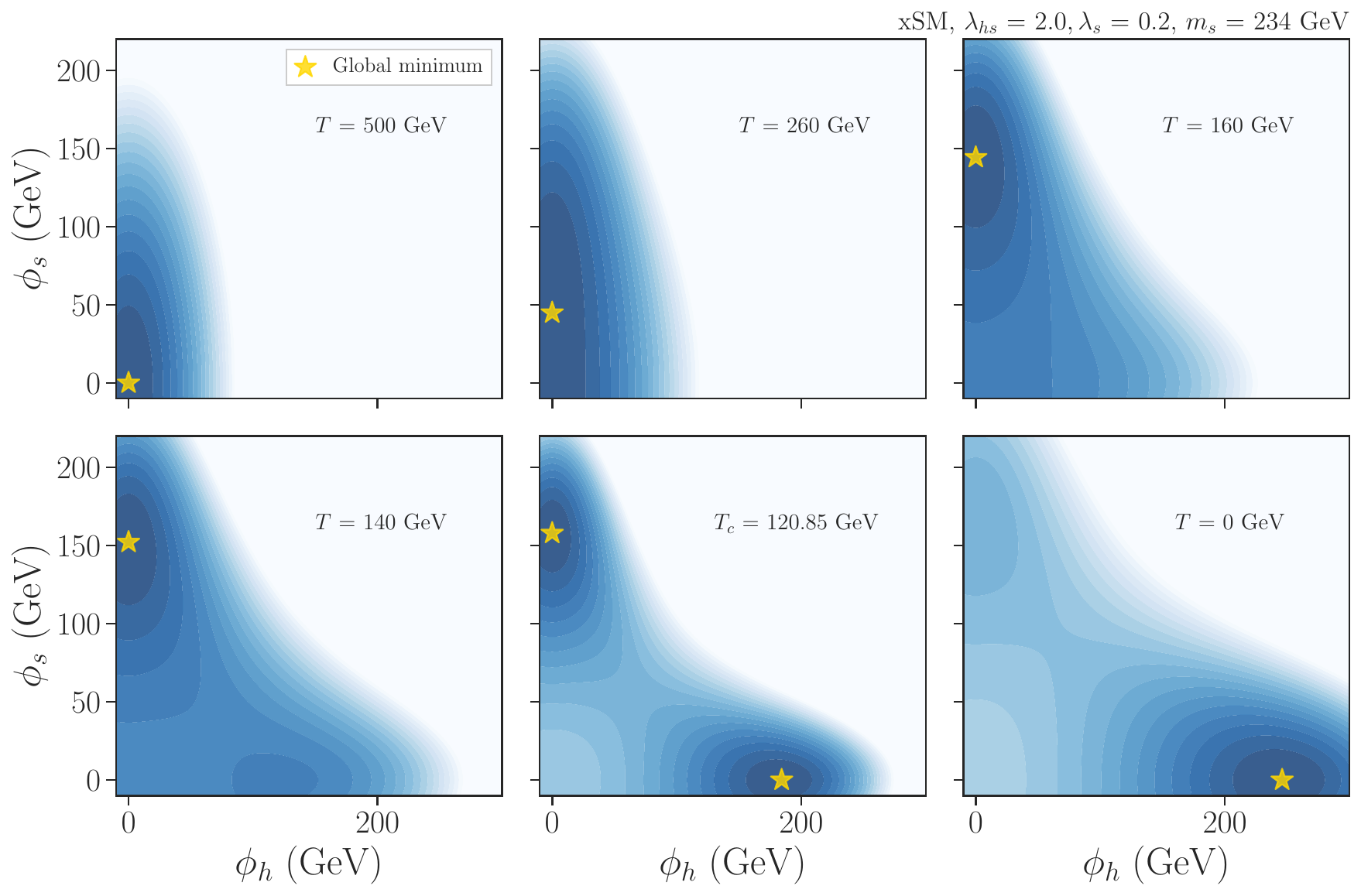}
\caption{Typical evolution history of the thermal effective potential of the xSM. The color represents the magnitude of the effective potential, with darker colors indicating a lower potential. The global minimum of the potential is indicated by a yellow star. }
\label{fig:bk4}
\end{figure*}

\section{Result and Discussion}

The typical thermal history of the effective potential of the xSM can be divided into three stages, as shown in Fig.~\ref{fig:bk4}.  In the very early universe, at high temperature, symmetry is restored and the vacuum is located at the origin at $(\phi_h,\phi_s) = (0,0)$.  As the universe cools, $\phi_s$ smoothly develops a non-zero expectation value, so the vacuum becomes $(0,v_s\neq 0)$.  With the temperature further dropping, a new minimum of $(v_h\neq0,0)$ appears, which becomes the desired EW-breaking vacuum $(v_h=246~\gev,v_s=0)$ when the temperature goes to zero. 
This thermal evolution of the potential allows for three possible transition scenarios.  The universe could tunnel from $(0,v_s)$ to $(v_h,0)$, and so the electroweak symmetry is successfully and correctly broken. 
The other two scenarios have been shown in Fig.~\ref{fig:bk1}, where the tunneling probability from $(0,v_s)$ into $(v_h,0)$ is too low and the $V(0,v_s)$ always smaller than $V(v_h,0)$. 
These two scenarios are generated using two benchmark points of the xSM, which have $m_{s} = 234$ and $220~\gev$ respectively, together with fixed $\lambda_{hs}=2.0$ and $\lambda_{s} = 0.2$.  To visualize the two-dimensional potential in one-dimension, we only show the potential along the straight line between the two minima in the $(\phi_h,\phi_s)$ plane. 
In these two situations, checking whether the desired EW-breaking vacuum is stable or long-lived is meaningless. Consequently, in order to arrive to a reliable conclusion for vacuum stability, it is unavoidable to study the thermal history of the universe.

For the benchmark point in the left frame of Fig.~\ref{fig:bk1}, when $T>267.9~\gev$ there is only one minimum in the potential located at the origin. Below that temperature 
the singlet field develops a non-zero expectation value smoothly increasing from $0$ to $168.3~\gev$ until zero-temper
ature. A new minimum of $(v_h=54.7~\gev,v_s=0)$ appears at $T=147.6~\gev$ and evolves into the desired EW-breaking vacuum $(v_h=246~\gev,v_s=0)$ as the temperature goes to zero. The two minima are degenerate at $T_c=120.9~\gev$, so after that the universe may transition from $(0,v_s)$ to $(v_h,0)$. We find, however, that the maximal probability of non-transitioning $P(T_{opt},0)$ is approximately 100\%. Therefore, it is unlikely that the universe will evolve into the desired EW-breaking vacuum for this benchmark point. Only looking at the zero-temperature potential, however, we see a stable EW-breaking vacuum which has no stability problem. This demonstrates the importance of calculating transitions in the cosmological history.   

The benchmark point of the right frame of Fig.~\ref{fig:bk1} has similar minimum structure as above. The difference is that there is no critical temperature between the two minima $(0,v_s)$ and $(v_h,0)$, i.e. the free energy of $(0,v_s)$ is larger than that of $(v_h, 0)$ throughout. 
As a result, the universe always stays in the EW-restored minimum. In the zero-temperature potential, the desired EW-breaking vacuum is meta-stable with a lifetime much longer than the universe's age, so it is acceptable in the traditional method. For this kind of situation, checking the history is time-saving, because one only needs to trace the position and free energy of the minima, no transition calculation is needed. 

The above discussion shows that it is necessary to modify the procedure for accessing vacuum stability.  Below, we assume that this is the case for even more complicated models, and the universe is located at the EW-restored vacuum at some high temperature.  Then, we can assess vacuum stability in following steps staring from the EW-restored vacuum.
\begin{itemize}
\item[(1)] Map out the position and free energy of the minima of the effective potential as the function of temperature (e.g. using numerical tools~\cite{wainwright2012cosmotransitions,Athron:2020sbe}).
\item[(2)] As the temperature is lowered,
whenever degenerate minima occur, i.e. if there is a critical temperature $T_{c}$, calculate the transition probability and transition temperature below $T_c$. 
\item[(3)] If the transition probability is larger than a certain threshold, set the vacuum of the universe to be the new minimum, and go to Step (2) until $T=0$. 
\item[(4)] Check whether the vacuum of the universe at $T=0$ is the desired EW-breaking vacuum. If not, the given parameter point is excluded. If yes, it is allowed when the vacuum is stable or long-lived, and is excluded when the vacuum is short-lived.
\end{itemize} 

%%%fig.3 
\begin{figure}[t]
\centering 
\includegraphics[width=0.48\textwidth]{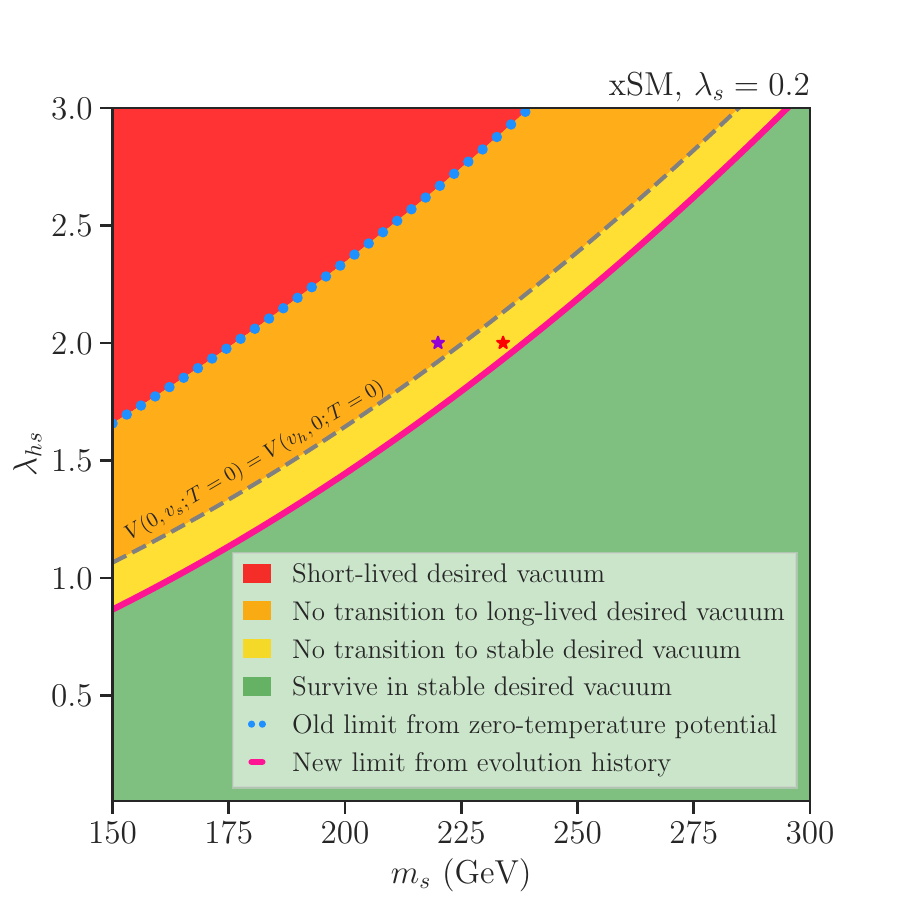}
\caption{Classifications of vacuum structure in parameter space of the xSM. In the red region, the desired EW-breaking vacuum is meta-stable with a life-time shorter than the age of the universe. In the orange region, the desired vacuum is long-lived but always has higher free energy than the EW-restored vacuum in the cosmological history. The desired vacuum in the yellow region is stable but the early universe can not transition to it. Only in the green region the universe can safely reach to the stable desired SM vacuum. 
Therefore, the dotted and solid curves represent respectively limits from checking minima in the zero-temperature potential and the evolution history of the universe. 
The red~(purple) star indicates the benchmark point shown in the left (right) panel of Fig.~\ref{fig:bk0} and Fig.~\ref{fig:bk1}. }

\label{fig:bk3}
\end{figure}

Following this procedure, we explore the parameter space of the xSM. For instance, the result for fixed $\lambda_{s} = 0.2$ is displayed in Fig.~\ref{fig:bk3}, where we compare the limit from only checking the low-temperature potential and the limit from considering the evolution history of the universe.  The parameter space is split by the dashed curve, at which the two minima $(v_h,0)$ and $(0,v_s)$ are degenerate at zero-temperature, and the parameters satisfy 
\begin{equation}
    \lambda_{hs} = \frac{2}{v_{h}^2} \left(m_{s}^2 + \mu_{h}^2\sqrt{\frac{\lambda_{s}}{\lambda_{h}}} \right)
\end{equation}
in the on-shell-like renormalization scheme. The red star in the yellow region and the purple star in the orange region correspond the situations described in the left and right frames of Fig.~\ref{fig:bk1}, respectively. For clarity, we list additional benchmark points selected from different regions in Tab.~\ref{tab1}. 

\begin{table*}[ht]
\centering
\caption{Benchmark points representing different regions in Fig.~\ref{fig:bk3} with fixed values of $\lambda_s=0.2$ and $\lambda_{hs}=2.0$. $\tau/\tau_{\rm uni}$ represents the ratio of the lifetime of the desired EW-breaking vacuum decaying to the EW-restored vacuum to the lifetime of the universe. $T_c$ and $T_n$ stand for the critical temperature and the nucleation temperature, respectively, of the transition from the EW-restored vacuum to the EW-breaking vacuum.
`-' indicates that there is no corresponding vacuum or quantity for the benchmark point.
The units of all quantities with mass dimensions are \gev. } 
\label{tab1}
\begin{tabular}{c|c|cc|cc|cccc}
\hline
\hline
 & Color of & \multicolumn{2}{c|}{EW-restored vacuum at $T=0$} & \multicolumn{2}{c|}{EW-breaking vacuum at $T=0$}  & $\tau/\tau_{\rm uni}$ &  $T_c$ & $T_n$\\
$m_s$ & the region & ~$(\phi_h,~\phi_s)$ & $V_{\rm eff}$ & ~$(\phi_h,~\phi_s)$ & $V_{\rm eff}$ \\
\hline
150 & red & (0,~427) & $-1.76 \times 10^9$ & (246,~0) & $-1.26 \times 10^8$ & $2.04 \times 10^{-28}$ & - & - \\
179 & orange & (0,~374) & $-1.01 \times 10^9$ & (246,~0)  & $-1.25 \times 10^8$ & 4.96 & - & - \\
220 & orange & (0,~246) & $-1.81 \times 10^8$ & (246,~0)  & $-1.24 \times 10^8$ & $\infty$ & - & - \\
234 & yellow & (0,~168) & $-3.93 \times 10^7$ & (246,~0) & $-1.23 \times 10^8$ & - & 120.8 &  - \\
241 & green & (0,~106) & $-6.58 \times 10^6$ & (246,~0) & $-1.23 \times 10^8$  & -  & 147.7 & 146.8\\
260 & green & - & - & (246,~0) & $-1.22 \times 10^8$ & - & 151.4 & 151.4 \\
\hline
\end{tabular}
\end{table*}

Above the dashed curve, the EW-restored vacuum is the  global minimum at zero-temperature as well as at finite temperature, as demonstrated by the first three benchmark points in Tab.~\ref{tab1}. The energy gap between the two vacua increases with $\lambda_{hs}$ increasing, leading to an increase in the decaying lifetime. Without considering the cosmological history, the desired EW-breaking vacuum is meta-stable, but can be long-lived, such as in the orange region of Fig.~\ref{fig:bk3}. Therefore, the transitional limit is the upper bound of the orange region. Thermal history tells us that the universe starts from the EW-restored vacuum and remains there as it is global minimum throughout. So the orange region is actually ruled out. 

Below the dashed curve, the desired EW-breaking vacuum at zero-temperature is the global minimum and thus stable. In the yellow region, the energy gap between the vacua is not large enough to achieve the transition that breaks electroweak symmetry. 
This is why the fourth benchmark point in Tab.~\ref{tab1} has a critical temperature but no nucleation temperature. 
Only in the green region, the early universe transits from $(0,v_s)$ or $(0,0)$ to $(v_h,0)$ and then evolves to $(246~\gev,0)$ at $T=0$. 

There are two phase transition patterns in the green region. 
In the vicinity of the solid pink curve, the early universe vacuum undergoes transitions from $(0,0)$ to $(0,v_s)$ to $(v_h,0)$. The first one $(0,0) \to (0,v_s)$ is a smooth crossover transition typically occurring at around $150\sim 200$ GeV. The subsequent step $(0,v_s) \to (v_h,0)$ is a first-order phase transition at $100\sim 150$ GeV. If this transition is sufficiently strong, it may lead to detectable stochastic gravitational waves, as explored in Ref.~\cite{Vaskonen:2016yiu,Beniwal:2017eik}. The duration of this phase transition is typically short. For example, the penultimate benchmark point in Tab.~\ref{tab1} has $T_c=147.7\gev$ and $T_n=146.8\gev$, indicating a rapid transition.
In the case of extreme supercooling, the phase transition could be delayed until the MeV  scale~\cite{Xiao:2023dbb}. However, achieving this requires highly fine-tuned parameters near the pink solid curve.The effects of extreme supercooling on the results are minimal.
In the region well below the pink solid curve, the early universe vacuum transitions directly from $(0,0)$ to $(v_h,0)$ through a crossover transition, similar to the SM. Consequently, no matter-antimatter asymmetry or gravitational waves are produced.

The $\lambda_{hs}$ value of the new limit~(pink solid curve) is about 0.2 and 0.9 lower than that of the dashed curve and the old limit~(blue dotted curve). This new limit is consistent with previous studies of xSM, such as Refs.~\cite{Beniwal:2017eik, Kurup:2017dzf}, and our main purpose here is to demonstrate that it is more stringent than the limit from the traditional vacuum stability analysis. 

We see that the traditional method accepts some regions of meta-stable EW-breaking vacuum, but checking the cosmological history proves that incorrect. The exclusion of such regions is computationally fast, because one only needs to trace the minima of the potential with temperature. On the other hard, it is more time-consuming to exclude the yellow region, as it involves calculation of the transition probability at finite temperature. 

The improvement of the new limit on vacuum stability is significant in the xSM, let alone in more complicated models. With more than one first-order phase transitions in the thermal history, such as in a supersymmetric model~\cite{Athron:2019teq}, every transition has to be examined carefully. The most popular vacuum stability limits in the Minimal Supersymmetric Standard Model (MSSM) are derived to prevent the desired EW-breaking vacuum from tunneling into the color-breaking vacuum. These limits result in constraints on stop and stau sectors, such as~\cite{Blinov:2013fta, Duan:2018cgb}
\begin{equation}
    A_t^2 < 3.4(m^2_{Q_{3,3}}+m^2_{u_{3,3}})+60(m^2_{H_u}+\mu^2)
\end{equation}
Here thermal corrections to the effective potential are included, but not from the view of tracing cosmological history. There is a possibility that the universe underwent a phase transition into the color-breaking vacuum or another vacuum state during its early stages of evolution, which give a new limit. 
The relationship between the old and new limits may be tangled, and deserves case by case investigation. 

%\indent
%The more stringent constraints from the vacuum stability considering the thermal history have important implications. For example, some predictions for the signal rates of N2HDM at the LHC may be excluded by such stringent constraints, which implies that the detailed thermal history in a given new physics model may be glimpsed through the LHC experiments~\cite{biekotter2021fate}. Moreover, these constraints can also affect the electroweak baryogenesis. In baryogenesis, a strong first-order phase transition is one of the vital ingredients and is often characterized as
%\begin{equation}
%    \frac{v_c}{T_c} \ge 1,
%\end{equation}
%where $v_c$ is the vacuum expectation value of the SM vacuum at the critical temperature. However, there exist some points that meet the condition but cannot reach the SM vacuum at zero-temperature~\cite{2hdmtrapped}. Thus these points are nonphysical and should be excluded. 
%\\
%\indent The xSM model is so simple that we can trace the thermal history easily and analyze its impact thoroughly. Although this simple model may not describe the real new physics, it can serve as an example. The study in \cite{cline1999electroweak} pointed out that in the absence of squark mixing and assuming a heavy $A_0$ boson, the effective potential of the MSSM is the same as the xSM. So we can study the vacuum problem (say charge or color breaking ) of the MSSM in a similar way. 

\section{Conclusions}
\label{sec:conclusions}
We argue that the study of vacuum stability should involve full consideration of the cosmological history, which is typically ignored in this context.  With the experimental discovery of gravitational waves, the impact of BSM physics on the cosmological history draws attention because a strong first order electroweak phase transition may generate detectable gravitational wave signals.  Our study shows that there are parameter space regions where the transition to the desired EW-breaking vacuum is of low probability or does not even happen, which is not taken into account by the traditional vacuum stability analysis. 

In this paper we studied vacuum stability taking into account the thermal history of the universe. After demonstrating the lack of transition to the desired EW-breaking vacuum in some parameter points of the xSM, we gave a general procedure which incorporates cosmological history in the vacuum stability study, and compared the traditional and new limits.  It is worth noting that this limit is an extension of the vacuum stability limit, which needs to be considered even if baryogenesis is not considered.  We found that checking the cosmological history can provide a much stringent and sometimes computationally cheaper limit of vacuum stability.  In turn, this motivates further studies of phase transitions in new physics models.

\section*{Acknowledgements}
This work was supported by the National Natural Science Foundation of China (2105248, 11821505, 12075300), the Australian Research Council (DP180102209, DP210101636),  Peng-Huan-Wu Theoretical Physics Innovation Center (12047503),
Key R\&D Program of Ministry of Science and Technology (2017YFA0402204), and by the Key Research Program of the Chinese Academy of Sciences (XDPB15). CB is indebted to Peter Athron for early contributions developing the 
idea of using the cosmological phase history leading to the EW vacuum as 
a new constraint and to Lachlan Morris for work realising this with CB 
and Peter Athron in another project parallel to this one.

\bibliographystyle{elsarticle-num} 
\bibliography{bibliography}

%%\end{thebibliography}

\end{document}